\documentclass[prd,superscriptaddress,showpacs,twocolumn,nofootinbib,floatfix,amsmath,amssymb]{revtex4-1}
\usepackage{mathtools}
\usepackage[usenames,dvipsnames]{xcolor}
\usepackage{color}
\usepackage{braket}
\usepackage{bm}
\usepackage{graphicx}
\usepackage{hyperref}
\usepackage{epstopdf}

\def\mat#1{\bm{\mathrm{#1}}}
\def\op#1{\hat{#1}}

\newcommand{\ii}{\mathrm{i}}

\begin{document}

\author{Krzysztof Lorek}
\affiliation{Institute of Theoretical Physics, Faculty of Physics, University of Warsaw, Pasteura 5, 02-093 Warsaw, Poland}

\author{Daniel Pecak}
\affiliation{Institute of Theoretical Physics, Faculty of Physics, University of Warsaw, Pasteura 5, 02-093 Warsaw, Poland}

\title{Extraction of genuine tripartite entanglement from the vacuum}
\author{Eric G. Brown}
\email{e9brown@uwaterloo.ca}
\affiliation{Department of Physics and Astronomy, University of Waterloo, Waterloo, Ontario N2L 3G1, Canada}

\author{Andrzej Dragan}
\email{dragan@fuw.edu.pl}
\affiliation{Institute of Theoretical Physics, Faculty of Physics, University of Warsaw, Pasteura 5, 02-093 Warsaw, Poland}

\begin{abstract}
We demonstrate and characterize the extraction of genuine tripartite entanglement from the vacuum of a periodic cavity field. That is, three probe quantum systems (detectors) can become both bipartitely and tripartitely entangled without coming into causal contact, by means of interaction with a common quantum field. We do this by using an oscillator-detector model that forgoes the need for perturbation theory and which instead is solved exactly. We find that the extraction of tripartite entanglement is considerably easier than that of bipartite. As a secondary result, we also compare a periodic cavity with one that has Dirichlet boundary conditions. We find that the extraction of both bipartite and tripartite entanglement is more easily achieved using the former case.
\end{abstract}

\maketitle

\section{Introduction}

It is known that the vacuum state of a relativistic quantum field is entangled. The entanglement structure can be seen explicitly by describing the vacuum state from a doubly-uniformly accelerated Rindler frame (see for example \cite{Dragan13}), but it can also be a subject of interactions with external localized systems such as point-like Unruh-DeWitt detectors \cite{BirrellDavies}. In this case the interactions lead to entanglement swapping from the field onto the pairs of detectors that have never been in contact in the past \cite{Reznik03}. These detectors (for example qubits or harmonic oscillators) can then potentially be used for quantum computational purposes. That is, we are extracting a resource (entanglement) from a reservoir.
So far the studies have largely been limited to the case of bipartite entanglement because perturbative calculations involving more than two detectors are quite intense and because, in general, computing tripartite entanglement measures for mixed states can be exceedingly difficult. In this paper we demonstrate the existence of tripartite entanglement in a vacuum field and characterize its extraction by a trio of Unruh-DeWitt-type detectors. One may 
find this surprising given the fact that all three-point correlation functions vanish in the vacuum. However while it is true that GHZ-type tripartite entanglement is witnessed by the three-point function, tripartitely entangled states also include those of W-type entanglement, which may be non-zero even when the three-point function is vanishing. Furthermore we show that, surprisingly, in the case under study the tripartite entanglement turns out to be more easily accessible than the bipartite. We observe a rich and interesting structure of the vacuum entanglement that has yet to be fully understood.

To probe and harvest the vacuum entanglement we employ the Unruh-DeWitt detector model with the two-level systems replaced by harmonic oscillators. Since the vacuum state is Gaussian and the interaction Hamiltonian quadratic --- the final state of the overall system will remain Gaussian. This approach allows one to perform analytical calculations without the need of using perturbative expansion, as proposed in \cite{Dragan11}. The authors, however, took into account only a single mode of the field that lead to problems with causality violation. The problem was fixed in \cite{Brown13} by adding more field modes and the idea was further developed, providing a useful formalism for the analytical treatment of point-like detectors interacting with the vacuum. This formalism has been extensively used to study various scenarios in the harvesting of entanglement \cite{Dragan11,Brown13,Martinez13} and more general quantum correlations \cite{Brown13(2)} in the case of bipartite harvesting (i.e. two 
detectors).

Here we employ this approach for the case when three point-like detectors, initially all uncorrelated in and their ground states, interact with the vacuum state of the scalar massless field. We only consider the case when the time of interaction is shorter than the light crossing time between the detectors. This rules out the possibility of creation of entanglement via direct interactions between detectors. All of the entanglement generated between the detectors must therefore have been harvested from the pre-existing entanglement present in the vacuum. In particular we consider the case of periodic field in which the three detectors are placed equidistantly. Although the final state of the detectors alone is mixed, this symmetry allows us for the easy verification of genuine tripartite entanglement between them that has been swapped from the field. We compare the regime of parameters for which the tripartite entanglement is harvested with the parameters for which bipartite entanglement is 
achieved between two detectors alone, and we find that in fact tripartite entanglement is more easily harvested.

In addition, we note that both bipartite and tripartite
entanglement are considerably more easily harvested
within a periodically-identified cavity, as opposed to one
with Dirichlet boundary conditions (i.e. mirrors). This
fact may be important knowledge, for example, when designing
an experimental setup for entanglement harvesting.


It should be noted that a previous paper, \cite{Silman05}, has also demonstrated the extraction of W-type tripartite entanglement from a quantum field. The differences between this work and our own are significant however, in that here our computations are entirely non-perturbative and we provide detailed data on the regions in parameter space in which the detectors obtain tripartite entanglement. Furthermore in our study we consider a cavity field rather than one in free space. This is important because any verification and utilization of such harvested entanglement is very likely to come about through a cavity-type scenario similarly as in the case of \cite{Delsing}.

In Sect.~\ref{setting} we present the scheme we are using for the calculations, in Sect.~\ref{results} we show the results and compare different settings and finally Sect.\ref{conclusions} concludes the paper.

In all the calculations we use natural units such that $c=\hbar=1$.

\section{The setting and model \label{setting}}
In this work we consider three detectors in a one-dimensional cavity of length $L$ with a massless scalar field. Both the field and the detectors are initially in their ground states and are thus not entangled with each other. At time $t=0$ the detectors start interacting with the field and the system is allowed to evolve until time $t=T$. After this interaction the entanglement between the detectors is examined.

The detectors are modeled by harmonic oscillators with corresponding annihilation operators labeled by $\hat{d}_j$, $j\in\{1,2,3\}$ and the interaction between the detector and the field is of the type of Unruh-DeWitt. For more in-depth information on the oscillator-detector model the reader is referred to \cite{Brown13}. The formalism also involves the use of Gaussian quantum mechanics; the unfamiliar reader may find many review articles on this subject, for example \cite{Adesso07}.

\subsection{Periodic boundary conditions}
We consider two types of cavity walls boundary conditions. The first type is periodic boundary condition, for which the cavity field operator has the following expansion into the frequency modes:
\begin{equation}  \label{PerPhi}
\hat{\phi}(x)=\sum_{n=-\infty}^{\infty} \frac{1}{\sqrt{4\pi|n|}} (\op a_n e^{\ii k_n x}+ \op a_n^\dagger e^{-\ii k_n x}),
\end{equation}
where $k_n=\frac{2\pi n}{L}$. The above sum excludes the zero-mode ($n=0$). The Hamiltonian governing the free evolution of the system is given by:
\begin{equation}  \label{PerHfree}
\hat{H}_\text{free}=\sum_{j=1}^3\Omega\hat{d}_{j}^\dagger\hat{d}_{j}+\sum_{n}\omega_n\hat{a}_{n}^\dagger\hat{a}_{n},
\end{equation}
where the former term corresponds to the detectors and the latter to the field. Here $\Omega$ is the frequency of a detector, taken to be the same for each of the detectors; and $\omega_n=|k_n|$ (the field is massless).

The interaction Hamiltonian is in turn given by the standard Unruh-DeWitt interaction:
\begin{equation}  \label{PerHint}
\hat{H}_{int}=\sum_{j=1}^3\lambda(\hat{d}_{j}+\hat{d}_{j}^\dagger)\hat{\phi}(x_j)
\end{equation}
where $x_j$ is the position of the j-th detector, and $\lambda$ is the coupling, again taken to be the same for each of the detectors. Throughout the time of the interaction between $t=0$ and $t=T$ it has a constant finite value and it is zero otherwise.

\begin{figure}[]
    \centering
    \includegraphics[width=0.45\textwidth]{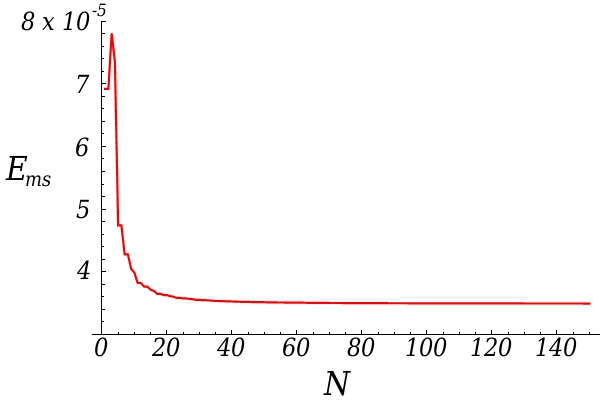}
    \caption{\small{(Color online)
             Convergence of the value of entanglement $E_{ms}$ (notation explained     
             in Sect.~\ref{results}) under increasing number of cavity field modes $N$. The parameter 
             values are: $\lambda=0.01$, $L=10$, $T=r$, $\Omega=0.4\pi$ in natural units and the 
             boundary conditions are Dirichlet.}}
    \label{convergence}
\end{figure}

In theory each detector interacts with the infinite number of cavity frequency modes. However in practice the modes of sufficiently high energy have negligible impact on the evolution of the detectors, and similarly the detectors have negligible impact on the evolution of these modes. We are thus safe in applying a UV cutoff to the field and only considering the evolution of the detectors with some finite number $N$ of field modes. In our work we choose $N$ such that increasing it further does not noticeably change the dynamics of the detectors. The results that we obtain, therefore, are equivalent to those that would be obtained from a complete scalar field, without a cutoff. Quantitatively, for the given scenario, the value of $N=50$ is sufficient for our results to converge. An example of this behaviour has been displayed in Fig. \ref{convergence}.

In order to utilize the phase-space Gaussian techniques that make oscillator-detectors so powerful we will need to represent these Hamiltonians by phase-space matrices. This is done as follows. First, we define the canonical quadrature operators for a given mode $j$ as
\begin{align}  \label{PerQuad}
	\op q_j = \frac{1}{\sqrt{2}}(\op a_j+\op a_j^\dagger), \;\;\;\;\; \op p_j=\frac{\ii}{\sqrt{2}}(\op a_j^\dagger - \op a_j),
\end{align}
where these definitions hold for both the oscillators and the field modes. We then define a vector of operators consisting of these operators, where the first entries correspond to the detector operators and the rest to the field. In the case of periodic boundary conditions, it has the form:
\begin{equation}  \label{vector}
\hat{\mathbf{x}}=(\hat{q}_{d1},\hat{p}_{d1},...,
\hat{q}_{d3},\hat{p}_{d3},\hat{q}_{-\frac{N}{2}},\hat{p}_{-\frac{N}{2}},...,\hat{q}_{\frac{N}{2}},\hat{p}_{\frac{N}{2}})^T
\end{equation}

From here, any quadratic (in general even time-dependent) Hamiltonian can be written as:
\begin{equation}    \label{PerHxfx}
\hat{H}(t)=\hat{\mathbf{x}}^T\mathbf{F}(t)\hat{\mathbf{x}},
\end{equation}
where $\mat F$ is a phase-space matrix that fully characterizes the Hamiltonian. As we will see, it is only the symmetric part of this matrix that impacts the evolution of the system. We thus define $\mat F^\text{sym} \equiv \mat F + \mat F^T$. From the Hamiltonians above it is straightforward to see that under the periodic boundary conditions the symmetrized matrices corresponding to the free and interaction Hamiltonians, respectively, are
\begin{equation}  \label{PerFfree}
\mathbf{F}_\text{free}^\text{sym}=\textrm{diag}(\Omega,\Omega,\Omega,\Omega,\Omega,\Omega,\omega_{-\frac{N}{2}},\omega_{-\frac{N}{2}},
...,\omega_{\frac{N}{2}},\omega_{\frac{N}{2}}),
\end{equation}
where note that $\omega_n=|k_n|=\omega_{-n}$, and
\begin{equation} \label{PerFint}
\mathbf{F}_\text{int}^\text{sym}=2\lambda \left[ \begin{array}{cc}
              \mathbf{0}_6 & \mathbf{X} \\
              \mathbf{X}^T & \mathbf{0}_{2N} \end{array} \right],
\end{equation}
where $\mathbf{0}_n$ is an $n\times n$ zero matrix, and the matrix $\mat X$ takes the form
\begin{widetext} 
\begin{eqnarray}   
	\mat{X} \equiv
	\begin{pmatrix} 
		\frac{\cos\left( k_{-N/2}x_1\right)}{\sqrt{2 \pi N}} & \frac{-\sin( k_{-N/2}x_1)}{\sqrt{2\pi N}} & \frac{\cos (k_{1-N/2} x_1)}{\sqrt{2 \pi(N-2)}} & \frac{-\sin (k_{1-N/2}x_1)}{\sqrt{2 \pi (N-2)}} & \dots & \frac{\cos (k_{N/2} x_1)}{\sqrt{2\pi N}} & \frac{-\sin (k_{N/2} x_1)}{\sqrt{2 \pi N}} \\
		0 & 0 & 0 & 0 & \dots & 0 & 0 \\
		\frac{\cos\left( k_{-N/2}x_2\right)}{\sqrt{2 \pi N}} & \frac{-\sin( k_{-N/2}x_2)}{\sqrt{2\pi N}} & \frac{\cos (k_{1-N/2} x_2)}{\sqrt{2 \pi(N-2)}} & \frac{-\sin (k_{1-N/2}x_2)}{\sqrt{2 \pi (N-2)}} & \dots & \frac{\cos (k_{N/2} x_2)}{\sqrt{2\pi N}} & \frac{-\sin (k_{N/2} x_2)}{\sqrt{2 \pi N}} \\
		0 & 0 & 0 & 0 & \dots & 0 & 0  \\
	\frac{\cos\left( k_{-N/2}x_3\right)}{\sqrt{2 \pi N}} & \frac{-\sin( k_{-N/2}x_3)}{\sqrt{2\pi N}} & \frac{\cos (k_{1-N/2} x_3)}{\sqrt{2 \pi(N-2)}} & \frac{-\sin (k_{1-N/2}x_3)}{\sqrt{2 \pi (N-2)}} & \dots & \frac{\cos (k_{N/2} x_3)}{\sqrt{2\pi N}} & \frac{-\sin (k_{N/2} x_3)}{\sqrt{2 \pi N}} \\
		0 & 0 & 0 & 0 & \dots & 0 & 0
	\end{pmatrix}.
\end{eqnarray}
\end{widetext}

\subsection{Dirichlet boundary conditions}
The second type of boundary condition that we will investigate is the the hard-wall Dirichlet boundary condition. For this case, the expansion into the frequency modes takes the form:
\begin{equation}  \label{DirPhi}
\hat{\phi}(x)=\sum_{n=1}^{\infty} \frac{1}{\sqrt{\pi n}} (\op a_n + \op a_n^\dagger) \sin(k_n x),
\end{equation}
where $k_n=\frac{\pi n}{L}$.

The description given above for the periodic boundary conditions, holds here as well. The Eqs. (\ref{PerHfree},\ref{PerHint},\ref{PerQuad},\ref{PerHxfx},\ref{PerFint}) are obeyed. There are however
the following differences:

Firstly due to a different mode structure, the vector (\ref{vector}) should be written here as:
\begin{equation} \label{vectordir}
\hat{\mathbf{x}}=(\hat{q}_{d1},\hat{p}_{d1},...,
\hat{q}_{d3},\hat{p}_{d3},\hat{q}_{1},\hat{p}_{1},...,\hat{q}_{N},\hat{p}_{N})^T
\end{equation} 

Hence also the form of the phase space free Hamiltonian is changed to:
\begin{equation}  \label{DirFfree}
\mathbf{F}_\text{free}^\text{sym}=\textrm{diag}(\Omega,\Omega,\Omega,\Omega,\Omega,\Omega,\omega_1,\omega_1,
...,\omega_N,\omega_N),
\end{equation}

While the phase space interaction Hamiltonian still has the block form given by (\ref{PerFint}), the matrix $\mat X$ now looks as follows:
\begin{eqnarray}   
	\mat{X} \equiv
	\begin{pmatrix} 
		      \frac{\textrm{sin} k_1x_1}{\sqrt{\pi}} & 0 
              & \frac{\textrm{sin} k_2x_1}{\sqrt{2\pi}} & 0 & \dots 
              & \frac{\textrm{sin} k_Nx_1}{\sqrt{N\pi}} & 0 \\
              0 & 0 & 0 & 0 & \dots & 0 & 0 \\
              \frac{\textrm{sin} k_1x_2}{\sqrt{\pi}} & 0 
              & \frac{\textrm{sin} k_2x_2}{\sqrt{2\pi}} & 0 & \dots 
              & \frac{\textrm{sin} k_Nx_2}{\sqrt{N\pi}} & 0 \\
              0 & 0 & 0 & 0 & \dots & 0 & 0 \\
              \frac{\textrm{sin} k_1x_3}{\sqrt{\pi}} & 0 
              & \frac{\textrm{sin} k_2x_3}{\sqrt{2\pi}} & 0 & \dots 
              & \frac{\textrm{sin} k_Nx_3}{\sqrt{N\pi}} & 0 \\
              0 & 0 & 0 & 0 & \dots & 0 & 0
	\end{pmatrix}.
\end{eqnarray}

\subsection{Time evolution}

Here we very briefly recapitulate the time-evolution method presented in \cite{Brown13}.

When working with Gaussian states, the state is fully characterized by the covariance matrix $\mat \sigma$, the entries of which are given by
\begin{align}
	 \sigma_{ij}\equiv \braket{\op x_i \op x_j+\op x_j \op x_i},
\end{align}
where $\op x_j$ are the entries of the vector in Eq. (\ref{vectordir}) or (\ref{vector}). Note that we are assuming here Gaussian states of zero-mean; extending the formalism beyond this assumption is straightforward but unnecessary here.

A Gaussian state will remain Gaussian over the course of evolution only if the Hamiltonian generating the evolution is quadratic or less in the quadrature operators (note that both $\op H_\text{free}$ and $\op H_\text{int}$ are as such), and this evolution is represented on phase space by a symplectic matrix $\mat S$ such that the covariance matrix (i.e. the state) evolves from time $t=0$ according to
\begin{equation}   \label{covEvo}
\boldsymbol\sigma(t)=\boldsymbol S(t)\boldsymbol\sigma(0)\boldsymbol S(t)^T
\end{equation}
As pointed out in \cite{Brown13}, the evolution matrix obeys a Schr\"odinger-type equation:
\begin{align}  \label{EOM}
	\partial_t \mat S(t)=\mat \Omega \mat F^\text{sym}(t) \mat S(t),
\end{align}
where $\mat F^\text{sym}$ is the symmetrized matrix corresponding to whatever Hamiltonian describes the system and $\mat \Omega$ is the symplectic form, given by
\begin{align} \label{symform}
	\mat{\Omega}=\bigoplus_i
	\begin{pmatrix}
		0 & 1 \\
		-1 & 0
	\end{pmatrix},
\end{align}
where $i$ runs through all degrees of freedom (both detectors and field modes). Here we opt to work directly in the Schr\"odinger picture (although this formalism can also be performed in the interaction picture; see \cite{Brenna13} for example) and so we will take $\mat F^\text{sym}=\mat F^\text{sym}_\text{free}+ \mat F^\text{sym}_\text{int}$ as given by Eqs. (\ref{PerFfree},\ref{PerFint}).

In our scenario we choose to sharply switch on the interaction at time $t=0$ and keep the coupling constant $\lambda$ constant during the course of evolution. As such, during the course of evolution we have that the Hamiltonian matrix $\mat F^\text{sym}$ is independent of time. This allows us to trivially solve the equation of motion, (\ref{EOM}), with the solution
\begin{equation}
\boldsymbol S(t)=\mathrm{exp}(\mathbf{\Omega F}^\text{sym}t).
\end{equation}
That is, the exact (non-perturbative) evolution of the system is fully given by taking a matrix exponential. Once $\mat S(t)$ has been solved, the evolved state (covariance matrix) is simply given by Eq. (\ref{covEvo}).

We have not yet specified the initial covariance matrix $\mat \sigma(0)$ to be used in Eq. (\ref{covEvo}). Since we are starting the field off in its vacuum state and the three detectors each in their respective ground states, the corresponding initial covariance matrix is just the identity matrix: $\mat \sigma(0)=\mat I_6 \oplus \mat I_{2N}$. In this specific case, the covariance matrix of the system at time $t$ is therefore $\mat \sigma (t)=\mat S(t)\mat S(t)^T$.

Once we have $\mat \sigma(t)$, the covariance matrix of the three detectors is simply the upper-left $6\times 6$ block, which we will label $\mat \sigma_{123}$. It can be further 
decomposed into smaller block matrices:
\begin{equation}
\boldsymbol\sigma_{123}=\left[ \begin{array}{ccc}
              \boldsymbol\sigma_1 & \boldsymbol\gamma_{12} & \boldsymbol\gamma_{13} \\
              \boldsymbol\gamma_{12} & \boldsymbol\sigma_2 & \boldsymbol\gamma_{23} \\
              \boldsymbol\gamma_{13} & \boldsymbol\gamma_{23} & \boldsymbol\sigma_3 \end{array} \right].
\end{equation}
The diagonal blocks are the covariance matrices of corresponding detectors alone, and the off-diagonal blocks contain information about the correlations between the different detectors. It is this matrix from which we will extract information regarding the entanglement among the detectors.

\subsection{Detectors' alignment}
In order to properly compare the amount of entanglement under Dirichlet and periodic boundary conditions, the detectors have been chosen in both cases to be aligned as $(x_1,x_2,x_3)=(\frac{1}{6}L,\frac{1}{2}L,\frac{5}{6}L)$, where $0$ and $L$ are the coordinates of the walls of the cavity. 

For periodic boundary conditions this alignment is symmetric in the sense that each detector is placed at the distance $\frac{L}{3}$ away from either of the remaining two, and hence the state of the system is invariant under the exchange of any two detectors. This property will allow us to easy determine whether or not the trio are tripartitely entangled.

For Dirichlet boundary conditions this alignment keeps the middle detector at the distance \nolinebreak $\frac{L}{3}$ away from the ones on the sides. In this case also the exchange of the detectors on the sides leaves the state of the system unchanged, however the asymmetry under the exchange of the middle detector with one on a side is unavoidable. For this reason we do not attempt to compute the extracted tripartite entanglement in the case of a Dirichlet cavity, but rather only compare the extracted bipartite entanglement with that of the periodic cavity.

\subsection{Bipartite entanglement} \label{Bi}

Given a covariance matrix for three detectors, we can calculate the amount of bipartite entanglement between each pair. As a measure of the bipartite entanglement, we will choose the logarithmic negativity \cite{Adesso04}. In order to calculate the logarithmic negativity between the $i$-th and $j$-th detector, we construct the reduced covariance matrix
of this pair by selecting out the appropriate blocks of $\boldsymbol\sigma_{123}$:
\begin{equation}
\boldsymbol\sigma_{ij}=\left[ \begin{array}{cc}
              \boldsymbol\sigma_i & \boldsymbol\gamma_{ij} \\
              \boldsymbol\gamma_{ij}^T & \boldsymbol\sigma_j \end{array} \right]
\end{equation}

The logarithmic negativity between the
$i$-th and $j$-th detector is then given by \cite{Adesso04}:
\begin{equation}
E_{ij}=\textrm{max}(0,-\textrm{log}_2\tilde{\nu}_{-})
\end{equation}
where $\tilde{\nu}_{-}$ is calculated from:
\begin{equation}
2\tilde{\nu}^2_{-}=\tilde{\Delta}-\sqrt{\tilde{\Delta}^2-4\textrm{det}\boldsymbol\sigma_{ij}}
\end{equation}
where $\tilde{\Delta}=\textrm{det}\boldsymbol\sigma_i + \textrm{det}\boldsymbol\sigma_j
-2\textrm{det}\boldsymbol\gamma_{ij}$.

\subsection{Tripartite entanglement} \label{Tri}

By definition, a tripartite system $ijk$ contains genuine tripartite entanglement if the state is inseparable across all three of the possible bipartitions $i|jk$ \cite{Adesso07}. Here what we truly care about is the existence or absence of tripartite entanglement, and less so its actual value. Thus we estimate the amount of tripartite entanglement using the geometrical average of the logarithmic negativities across all the bipartitions:
\begin{equation}
\bar{E}_{ijk}=\sqrt[3]{E_{i|jk}E_{j|ik}E_{k|ij}}.
\end{equation}
This quantity does not constitute a proper entanglement measure, but certainly it provides a yes-or-no answer to whether or not there is tripartite entanglement in the system. In the case of the periodic cavity, which is the scenario in which we actually utilize this, the fact that the three detector state is symmetric under detector permutation means that to determine the presence of tripartite entanglement we need only consider a single bipartition, since the other two will be equivalent. 

In general, computing the bipartite entanglement in a mixed, $1+2$ mode Gaussian state is difficult. In the case of the periodic cavity, however, we can use the symmetry under detector-swap to compute it easily. The fully symmetric state in this case will be of the form
\begin{align}
	\mat \sigma_\text{periodic}=\begin{pmatrix}
		\mat \sigma_1 & \mat \gamma & \mat \gamma \\
		\mat \gamma & \mat \sigma_1 & \mat \gamma \\
		\mat \gamma & \mat \gamma &\mat \sigma_1
	\end{pmatrix}.
\end{align}
The entanglement across, say, the bipartition $12|3$ can be easily computed by using an example of what has been called unitary localization \cite{Adesso04(2),Serafini05}. To this end, we consider applying a beam-splitter operation to the $12$ mode subsystem, which globally is given by
\begin{equation}
\mathbf{S}_\text{BS}=\begin{pmatrix}
              \mat I/\sqrt{2} & - \mat I/\sqrt{2} & \mat 0  \\
              \mat I/\sqrt{2} &  \mat I/\sqrt{2} & \mat 0 \\
	\mat 0 & \mat 0 & \mat I
	\end{pmatrix}.
\end{equation}
Being a unitary operation within the subsystem $1$ and $2$, this does not affect the entanglement across the $12|3$ bipartition. Furthermore, with mode symmetry this operation is seen to isolate all correlations solely between modes $2$ and $3$:
\begin{align}
	\mat{S}_\text{BS} \mat \sigma_\text{periodic} \mat S_\text{BS}^T=\begin{pmatrix}
		\mat \sigma_1-\mat \gamma & \mat 0 & \mat 0 \\
		\mat 0 & \mat \sigma_1+\mat \gamma & \sqrt{2} \mat \gamma \\
		\mat 0 & \sqrt{2} \mat \gamma & \mat \sigma_1
	\end{pmatrix}.
\end{align}
From here, we can apply the formula in Sect. \ref{Bi} to compute the entanglement between modes $2$ and $3$ of this transformed state. This will be equivalent to the bipartite entanglement across the $12|3$ partition which, as stated above for the periodic case, is a necessary and sufficient indicator of genuine multipartite entanglement among all three detectors.

In the case of the Dirichlet cavity the asymmetry between the detectors makes it much harder to the check the bipartite entanglement across all three partitions, and thus to check for the presence of tripartite entanglement. We therefore forgo this calculation and focus on the periodic cavity when considering tripartite entanglement.

\section{RESULTS\label{results}}

In this section we present the results that have been obtained for the detector alignment described above. The amounts of entanglement between detectors have been computed for both types of boundary conditions. We will show two results. First, that for a periodic cavity one can acausally (i.e. without causal contact between detectors) harvest tripartite entanglement, and that this is in fact easier to do so than for bipartite. Second, we will compare the harvesting of bipartite entanglement between the cases of a period and a Dirichlet cavity, finding that it is considerably easier to harvest in the periodic case.

All the results have been evaluated at a fixed value of the coupling constant $\lambda=0.01$.

The time of interaction $T$ will be presented here in the units of $r$, where $r=\frac{L}{3}$ is the light-crossing time between neighbouring detectors. Hence for times $T>r$ neighbouring detectors are within a timelike regime, and for $T<r$ within a spacelike one, meaning that in the latter case they could not have been in causal contact and all the entanglement generated in the system must be due to extraction from the field.

\subsection{Periodic boundary conditions}

\begin{figure*}[]
    \centering
    \includegraphics[width=1.05\textwidth]{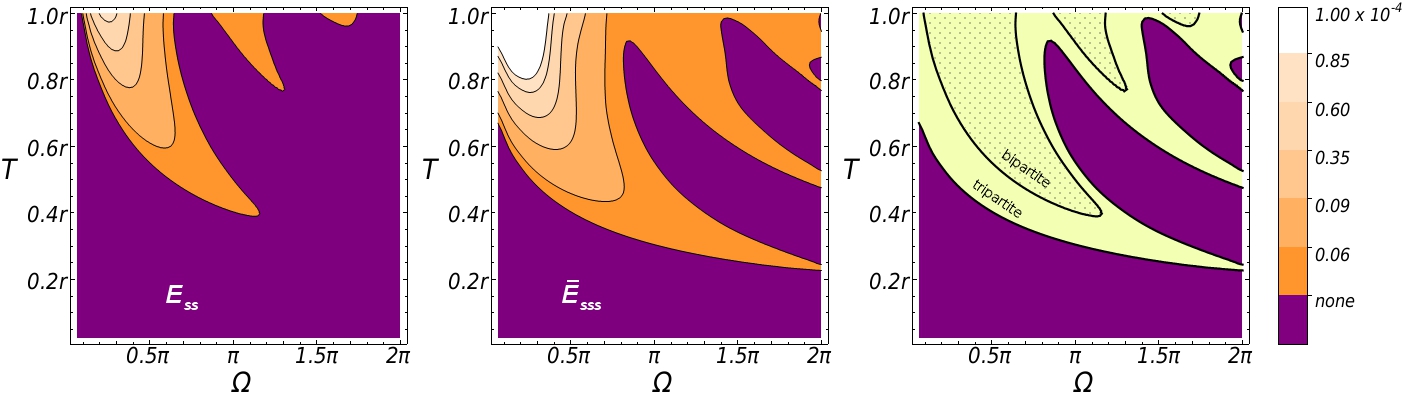}
    \caption{\small{(Color online) Entanglements $E_{ss}$, $\bar{E}_{sss}$ in the system subject to 
             periodic boundary conditions, as a function of $T$ and
             $\Omega$ at $L=10$ and $\lambda=0.01$; On the right: regions of existence
             of $E_{ss}$ and $\bar{E}_{sss}$ plotted together. All quantities are given in natural units.}}
    \label{periodic}
\end{figure*}

\begin{figure*}[]
    \centering
    \includegraphics[width=1.05\textwidth]{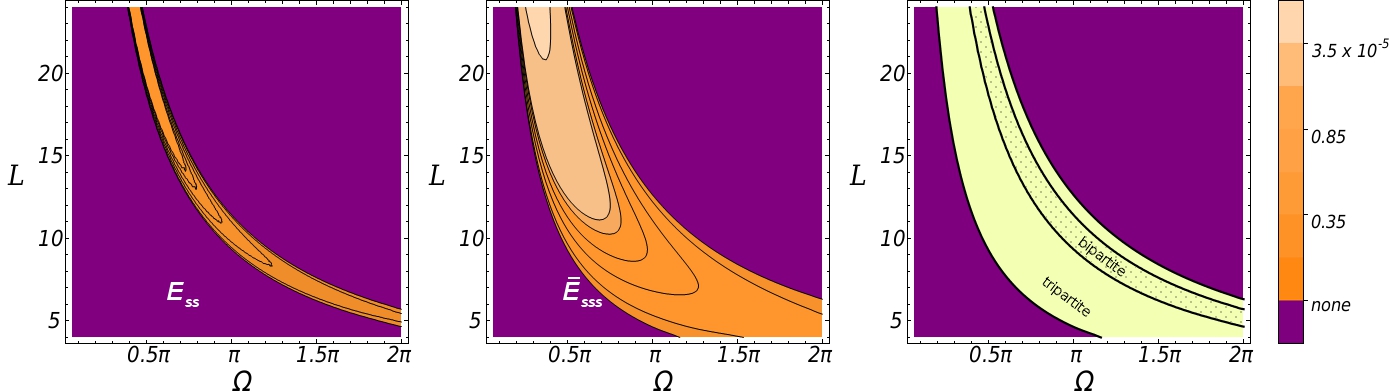}
    \caption{\small{(Color online) Entanglements $E_{ss}$, $\bar{E}_{sss}$ in the system subject to 
             periodic boundary conditions, as a function of $L$ and
             $\Omega$ at $T=0.4r$ and $\lambda=0.01$; On the right: regions of existence
            of $E_{ss}$ and $\bar{E}_{sss}$ plotted together. All quantities are given in natural units.}}
    \label{lo}
\end{figure*}

\begin{figure*}[]
    \centering
    \includegraphics[width=0.9\textwidth]{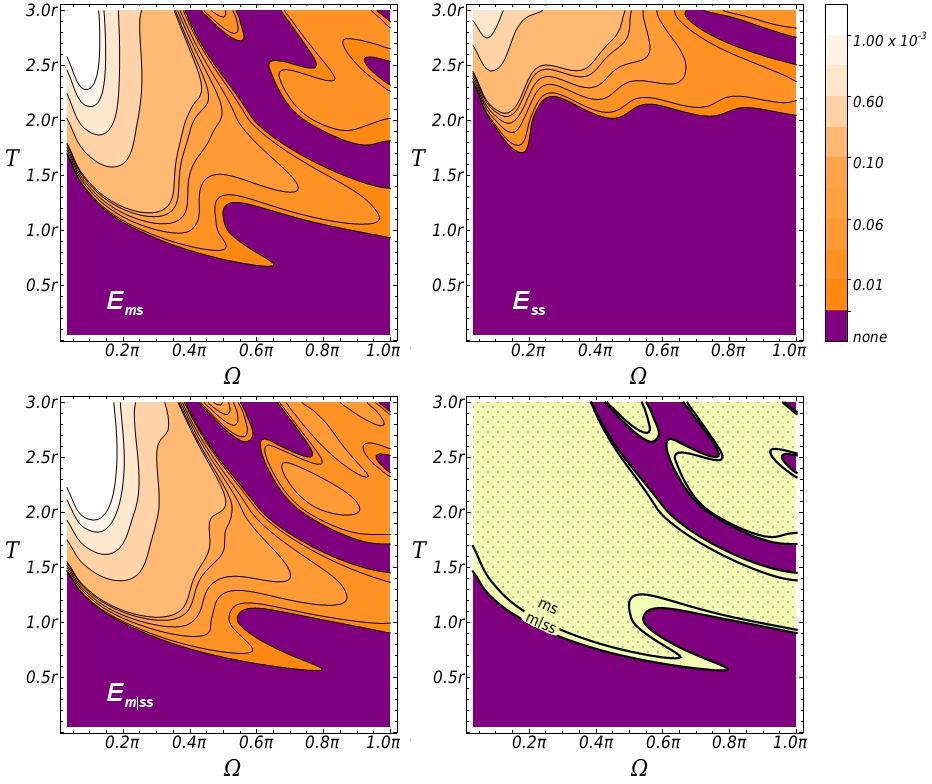}
    \caption{\small{(Color online) Entanglements $E_{ms}$, $E_{ss}$ and $E_{m|ss}$ in the system 
             subject to 
             hard wall boundary conditions, as a function of $T$ and
             $\Omega$ at $L=10$ and $\lambda=0.01$; Bottom right: regions of existence
             of $E_{ms}$ and $E_{m|ss}$ plotted together. All quantities are given in natural units.}}
    \label{dirichlet}
\end{figure*}

\begin{figure}[]
    \centering
    \includegraphics[width=0.45\textwidth]{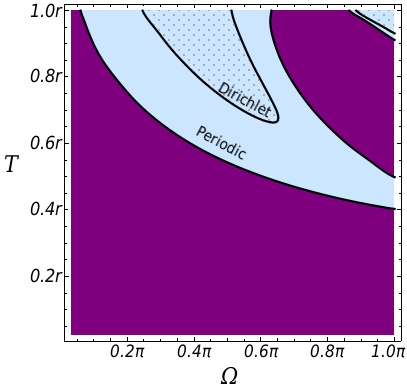}
    \caption{(Color online) A comparison, between the cases of period and Dirichlet boundary conditions, of the regions in which bipartite entanglement is harvested between neighboring detectors. Here $L=10$ and $\lambda=0.01$. All quantities are given in natural units.}
    \label{both}
\end{figure}

In this subsection due to symmetry we denote all the detectors by $s$, so $E_{ss}$
is the bipartite entanglement between a pair, which is identical for each pair. $\bar{E}_{sss}$ is the measure we use for tripartite entanglement,
which via equation (21) and symmetry is equal to $E_{s|ss}$. For simplicity the former will be referred
to simply as ``bipartite entanglement'' and the latter - as ``tripartite entanglement''.

For the case of periodic boundary conditions we have produced plots of the amount of entanglement as a function of $T$ and $\Omega$ (the frequency of the detectors), at a fixed value of $L=10$. This is given in Fig. \ref{periodic}, where we plot both $E_{ss}$ and $\bar{E}_{sss}$. In agreement with intuition, most entanglement is being produced after the light-crossing time of the neighbouring detectors, which is $r$. There are however regions in the $T,\Omega$ plane along which both types of entanglement persist
deeply into the spacelike regime. This entanglement does not diminish under increasing the number of field modes $N$, hence it is not an artifact of the imposed UV cutoff but rather a true physical effect. Moreover the regions where tripartite entanglement is 
produced are certainly broader than for those of bipartite, meaning that the
tripartite entanglement emerges earlier and is therefore easier to be harvested. This however is not surprising if we recall that $\bar{E}_{sss}=E_{s|ss}$ 
and we obtain $E_{ss}$ from $E_{s|ss}$
by tracing out one of the detectors. This is a local operation and so can only decrease the amount of entanglement, which implies that $E_{ss}$ always has to be less or equal $\bar{E}_{sss}$, as can be seen from our results.

To examine the dependence of the results on $L$ we have produced in Fig. \ref{lo} a plot of entanglement versus
$L$ and $\Omega$, having been produced at fixed $T=0.4r$. In the $L,\Omega$ plane we find a hyperbolic curve along which both bipartite and tripartite entanglement have been extracted by this time. Again, as must generally be the case, the region of non-zero tripartite entanglement is larger than that of bipartite. We also note that, within this scenario, the longer the cavity is and the smaller the detector frequency is the more entanglement can be acausally extracted.

\subsection{Dirichlet boundary conditions}

Throughout this subsection we denote the middle detector as $m$ and the ones on the sides as $s$.

The behaviour under Dirichlet boundary conditions has been found to be different to some extent. We have produced analogous plots of the amount of entanglement as a function of $T$ and $\Omega$, at fixed value of $L=10$. The comparison now involves the following quantities: the bipartite entanglement $E_{ms}$, the bipartite entanglement $E_{ss}$, and the bipartite entanglement $E_{m|ss}$ (which can be computed using the method of Sect. \ref{Tri}). We plot these in Fig. \ref{dirichlet}. Again we observe that most entanglement is produced after the corresponding light-crossing time, which for the case of neighbouring detectors is $r$, whereas for the ones on the sides is $2r$. Under these boundary conditions we have found points in the parameter space for which the entanglement $E_{ms}$, which is more broadly available, persists in the spacelike regime down to $T\sim 0.6r$. This is a considerably longer time than in the periodic case, which can be seen directly in Fig. \ref{both}. We clearly see that it is easier to extract bipartite entanglement in the periodic cavity than one with hard-boundary conditions.

The tripartite entanglement $\bar{E}_{mss}$ has not been calculated for this case, for reasons described in Sect. \ref{Tri}, however the bipartite entanglement $E_{m|ss}$ has been evaluated, which allows us to draw certain conclusions. The entanglement $E_{ms}$ and $E_{m|ss}$ have been plotted together for comparison and the region of existence of $E_{ms}$ is enclosed in the region of existence of $E_{m|ss}$, which is as it must be given the argument presented in the previous subsection. The region of non-zero $E_{m|ss}$ is slightly broader than that for $E_{ms}$ similarly to the periodic case.

The existence of $E_{m|ss}$ is a necessary (but not sufficient) condition for the existence of tripartite entanglement, hence we know that no tripartite entanglement can exist in the system if $E_{m|ss}=0$. We therefore conclude that, similar to bipartite entanglement, the extraction of tripartite entanglement is considerably easier to achieve using a periodic cavity rather than a Dirichlet one.

 Therefore we see from the plot that tripartite entanglement emerges no earlier than $T\sim 0.55r$. This is again considerably later than in the periodic case.


\section{CONCLUSIONS\label{conclusions}}

We have used non-perturbative Gaussian methods to examine the harvesting of both bipartite and of genuine tripartite entanglement from the vacuum field of a cavity. That is, a set of detectors interacting with a common quantum field can become entangled without causal contact by means of swapping the spatial entanglement present in the field.

There are two primary conclusions that we have made. First, from the vacuum state of a periodically-identified cavity field it appears that genuine tripartite entanglement can be harvested. This tripartite entanglement is expected to be of the W-type, due to the fact that the three-point functions of the vacuum field are vanishing. In fact, it is considerably easier to obtain tripartite entanglement than bipartite between any two of the three detectors. Indeed, we have been able to obtain tripartite entanglement after a time of interaction considerably smaller than the light-crossing time between pairs of detector. Specifically we see that a time as small as $t=0.21 r$, where $r$ is the distance between detectors, can be sufficient. We have provided detailed maps of the regions in parameter space in which bipartite and tripartite entanglement can be harvested.

Second, we have demonstrated that there is a significant difference between a periodic cavity field and one subject to Dirichlet boundary conditions (i.e. mirrors) for the harvesting of bipartite entanglement. Our finding is that such harvesting is considerably easier to achieve in a periodic cavity than in a Dirichlet one. Although we have not fully considered tripartite entanglement in the case of a Dirichlet cavity, the same conclusion can similarly be made for this.

Both of these results have applicability to the experimental confirmation as well as possible utilization of entanglement harvesting scenarios. They furthermore may be applicable to more general system-bath setups than what we have considered here, helping to point the way towards optimal strategies for generating quantum correlation utilizing such systems.


\acknowledgements
We thank A. Kempf and E. Mart\'{i}n-Mart\'{i}nez, as well as a special thanks to G. Adesso, for useful discussions.  E.G.B. gives his warm thanks to A.D. and the Institute of Theoretical Physics at the University of Warsaw for their hospitality during the visit in which this work was conceived. E.G.B. is further supported by the Natural Sciences and Engineering
Research Council of Canada.  A.~D. thanks for the financial support to National Science Center, Sonata BIS grant 2012/07/E/ST2/01402.


\begin{thebibliography}{10}

\bibitem{Dragan13}
A. Dragan, J. Doukas, E. Mart\'{i}n-Mart\'{i}nez, and D. E. Bruschi,
Class. Quantum Grav. {\bf 30}, 235006 (2013).

\bibitem{BirrellDavies}
N. D. Birrell and P. C. W. Davies,  {\em Quantum Fields In Curved Space} (Cambridge: Cambridge University Press, 1982).

\bibitem{Reznik03}
B. Reznik, Found. Phys. {\bf 33}, 167 (2003); B. Reznik, A. Retzker, and J. Silman, Phys. Rev. A {\bf 71}, 042104 (2005).

\bibitem{Dragan11}
A. Dragan, I. Fuentes, arXiv:1105.1192 [quant-ph] (2011).

\bibitem{Brown13}
E. G. Brown,  E. Mart\'{i}n-Mart\'{i}nez, N. C. Menicucci, and R. B. Mann,
Phys. Rev. D \textbf{87}, 084062 (2013).

\bibitem{Martinez13}
E. Mart\'{i}n-Mart\'{i}nez, E. G. Brown, W. Donnelly, and A. Kempf,
Phys. Rev. A \textbf{88}, 052310 (2013)

\bibitem{Brown13(2)}
E. G. Brown, Phys. Rev. A \textbf{88}, 062336 (2013)

\bibitem{Silman05}
J. Silman and B. Reznik, Phys. Rev. A \textbf{71}, 054301 (2005)

\bibitem{Delsing}
C. M. Wilson \textit{et al.}, Nature \textbf{479}, 376-379 (2011)

\bibitem{Brenna13}
W. G. Brenna, E. G. Brown, R. B. Mann, E. Mart\'{i}n-Mart\'{i}nez,
Phys. Rev. D \textbf{88}, 064031 (2013).

\bibitem{Adesso04}
G. Adesso, A. Serafini, and F. Illuminati, Phys. Rev. A {\bf 70}, 022318 (2004).

\bibitem{Adesso07}
G.~Adesso and F.~Illuminati,
\newblock J. Phys. A: Math. Theor {\bf 40}, 7821 (2007).

\bibitem{Adesso04(2)}
G. Adesso, A. Serafini, and F. Illuminati, Phys. Rev. Lett \textbf{93}, 220504 (2004)

\bibitem{Serafini05}
A. Serafini, G. Adesso, and F. Illuminati, Phys. Rev. A \textbf{71}, 032349 (2005)




\end{thebibliography}
\end{document}